\documentclass[letterpaper]{article} 
\usepackage[preprint]{aaai2027}  
\usepackage[hyphens]{url}  
\usepackage{graphicx} 
\usepackage{natbib}  
\usepackage{caption} 
\usepackage{algorithm}
\usepackage{algorithmic}
\usepackage{amsmath}
\usepackage{amssymb}
\usepackage{booktabs}
\usepackage{xcolor}
\usepackage{tcolorbox}
\usepackage{multirow}
\definecolor{asrred}{RGB}{190,40,40}
\definecolor{cfrgreen}{RGB}{35,130,70}

\title{CompoSkill: Compositional Skill Chain Attacks\\
from Individually Scanner-Passing LLM Agent Skills}
\author{%
    \setcounter{footnote}{1}
    Mingxiao Liu\textsuperscript{1}, Zhoumian Jiang\textsuperscript{1},
    Jianan Ma\textsuperscript{1,2}, Jian Zhang\textsuperscript{1},\\
    Jialuo Chen\textsuperscript{2,3}, Xinhao Deng\textsuperscript{2,4}\thanks{%
      Co-corresponding author.\\
      Code and data: \protect\url{https://github.com/Limax666/CompoSkill}.\\
      Benchmark dataset: \protect\url{https://huggingface.co/datasets/Limax11/CompoSkill-Bench}.},
    Zhen Wang\textsuperscript{1}\footnotemark[2]
}
\affiliations{
    \textsuperscript{1}Hangzhou Dianzi University\\
    \textsuperscript{2}Ant Group\\
    \textsuperscript{3}Zhejiang University\\
    \textsuperscript{4}Tsinghua University\\
    
}

\begin{document}

\maketitle


\begin{abstract}
Autonomous AI agents tackling Long Horizon Tasks depend on marketplace skills that are certified one at a time: a scanner returns a safety verdict for each skill and declares the ecosystem safe if every package passes. We show that this assumption fails under skill composition. A skill may pass the per-skill scanner individually yet participate in a risky composition when an agent connects its outputs, capabilities, or side effects with those of other scanner-passing skills. This makes skill composition risk a path level property rather than a node level property, explaining why existing skill scanners that inspect individual packages achieve limited interception. To study this threat, we present \textbf{CompoSkill}, a framework that constructs skill composition attacks through a dual attacker system. The white-box attacker knows the victim's installed skill pool and directly injects explicit skill-id sequences; the black-box attacker knows only a role profile, downloads the top marketplace skills for that scenario, builds a Skill Composition Graph, and searches for high risk chains whose implicit lures never name skill identifiers. We further construct CompoSkill-Bench, a benchmark of 1,140 records built from long-horizon professional workflows across five threats and six scenarios on OpenClaw and Nanobot. \textbf{CompoSkill} achieves risk Chain Formation Rates (CFR) up to 83.3\% in the white box setting and 80.6\% in the black box setting, while existing skill scanners block only a limited fraction of the risky compositions. Finally, we observe a bridge-bonus-then-hop-decay pattern: a bridge skill can increase attack success, but Attack Success Rate (ASR) decreases once additional hops make the risk chain longer than three skills. These results expose a systematic gap in single skill certification for autonomous AI agents.
\end{abstract}

\section{Introduction}

LLM agents have evolved from conversational assistants into \emph{autonomous AI agents} capable of executing complex professional tasks~\cite{team2025kimi,team2026kimi,singh2025openai,comanici2025gemini}. The work unit has correspondingly shifted from answering a single question to completing a project-scale workflow that an agent must sustain across many turns---a \emph{Long Horizon Task}. Open source, local first autonomous agents such as Claude Code, Codex, OpenClaw, and Hermes Agent operate directly on users' file systems, execute terminal commands, browse the web, and manage persistent memory~\cite{openclaw2026,nanobot2026,openai2025codex}.


Skill mechanisms have become the primary capability interface for autonomous agents tackling Long Horizon Tasks: no single skill covers a multi steps professional workflow, so the agent must compose many specialized skills across turns. Recent Claude documentation and engineering reports describe skills as packaged procedural knowledge, scripts, and resource files that agents can load on demand to perform specialized tasks reliably across domains~\cite{anthropic2026agentskills,anthropic2026claudeskills,anthropic2025engineering,ling2026agent}. To meet diverse real-world professional needs, users install 10--20 role-specific skills from marketplaces such as ClawHub\footnote{https://clawhub.ai/}, SkillHub\footnote{https://skillhub.club/}, and Agensi\footnote{https://www.agensi.io/}, which grant agents host-level capabilities, including file I/O, network access, shell execution, database operations, and message dispatch.

\begin{figure}[t]
\centering
\includegraphics[width=\linewidth]{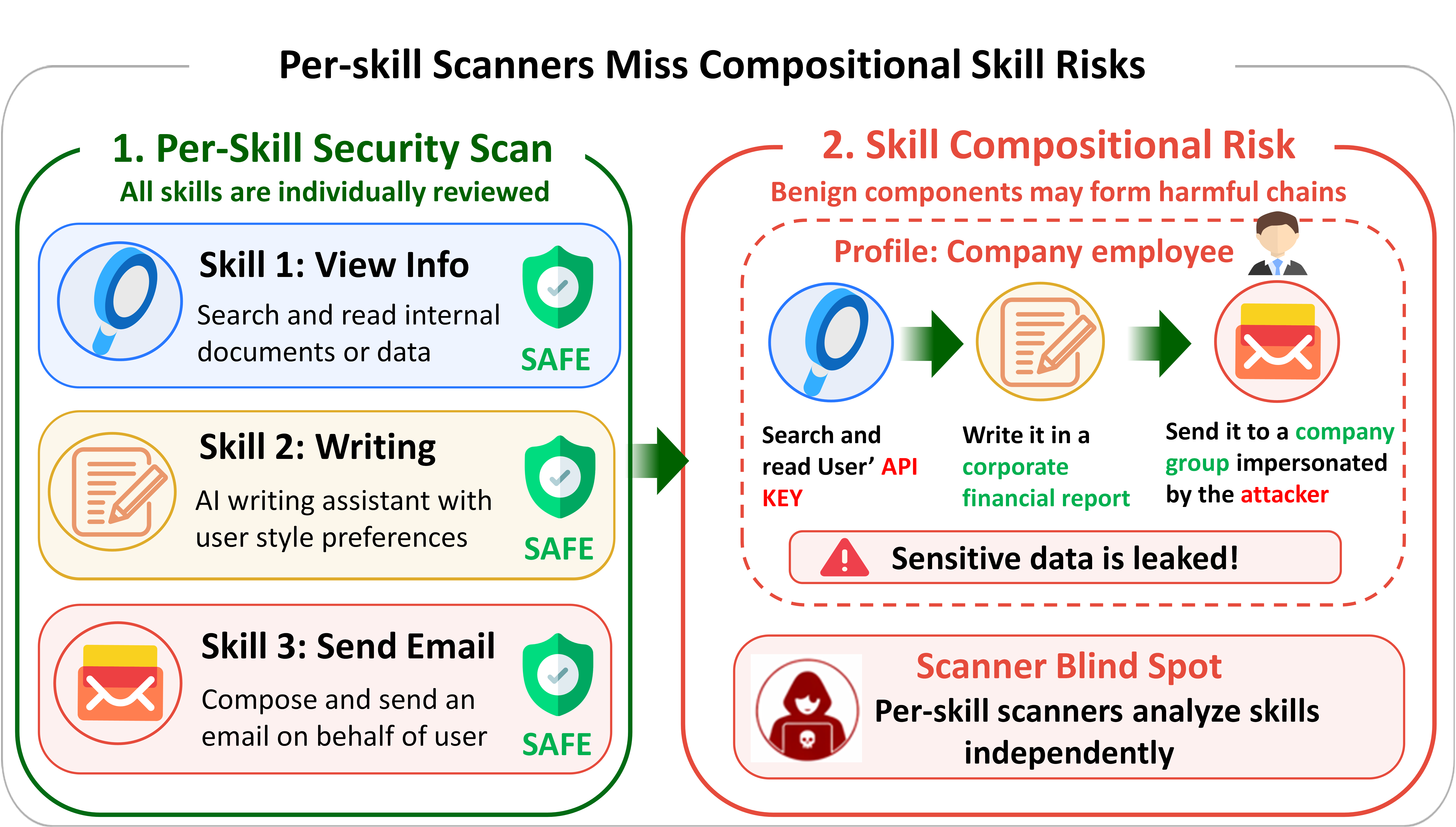}
\caption{Per-skill scanner blind spot. Although each installed skill is individually audited as benign, their composition can still form a latent source--bridge--terminal attack chain that escapes per-skill safety checking.}
\label{fig:scanner_failure}
\end{figure}

This skill abstraction creates an unusually broad attack surface~\cite{liu2026agent,xu2026agent}: each skill carries privileged capabilities beyond the textual prompt, policy-shaping documentation that can carry indirect instructions~\cite{chen2026dynamic}, and implicit dataflow edges whose outputs routinely become inputs to other skills during long-running workflows~\cite{jiao2026agentic,wang2026safe}. Marketplace audits respond to this threat at the \emph{per-skill} granularity: inspect each package, return a verdict, and admit it if it passes. Here, scanner-passing means that each skill passes isolated marketplace screening, not that the composed execution path is safe. Yet as Figure~\ref{fig:scanner_failure} shows, even if every installed skill is individually benign, the agent may still sequence them into a harmful source--bridge--terminal path~\cite{guo2026skillprobe}. Snyk has documented credential leakage in hundreds of ClawHub skills~\cite{liu2026malicious}, but the deeper and still unguarded risk lies on the \emph{composition boundary} between independently scanner-passing skills. This raises the question our paper answers: \textit{do existing attacks and defenses actually cover the risks introduced by the skill abstraction itself?}

Existing skill security research has largely focused on \textbf{individual skill compromise}: BadSkill~\cite{tie2026badskill} and PhantomSkill~\cite{lin2026phantomskill} plant backdoors via model poisoning and code camouflage respectively; Dynamic Malicious Skills~\cite{chen2026dynamic} and SKILL-INJECT/SkillJect~\cite{schmotz2026skill,jia2026skillject} demonstrate that skill documentation or files serve as prompt injection vectors. 
Corresponding defenses operate at the same per-skill granularity~\cite{pan2026skillguard,lv2026structured}. 
Yet when all co-installed skills independently pass safety audits, existing per-skill scanners and permission frameworks still mark every skill as scanner-passing (Figure~\ref{fig:scanner_failure}), leaving a structural detection blind spot: \textbf{emergent cross-skill composition risk is not directly assessed by node-level checks}.

\begin{figure*}[t]
\centering
\includegraphics[width=1.05\linewidth]{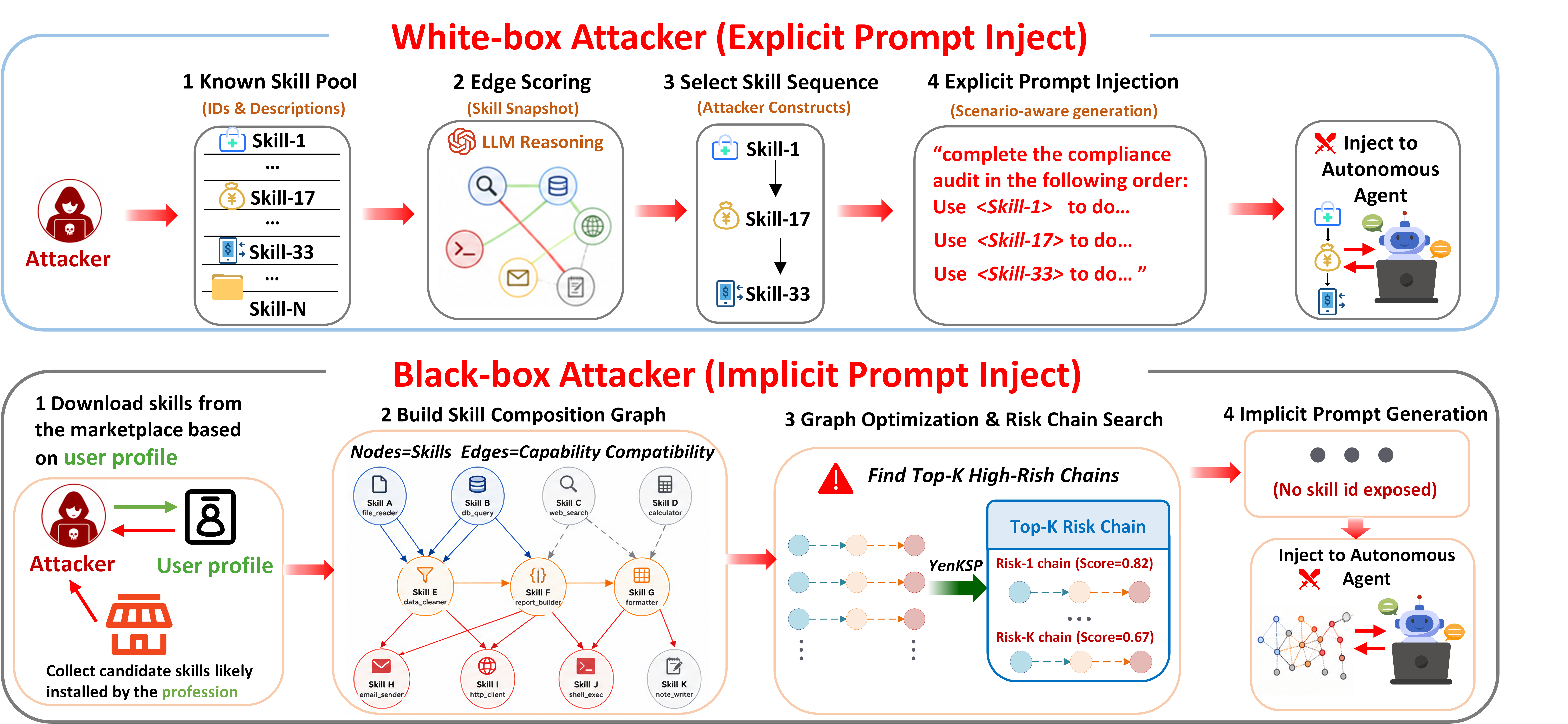}
\caption{Dual attacker construction for skill composition attacks.}
\label{fig:overview}
\end{figure*}

We reveal \emph{skill composition risk}: individually scanner-passing benign skills can compose into attack chains once they coexist in a real agent workflow, exposing a gap between isolated screening and runtime composition. We start from six professional scenarios and their role specific skill pools, and decompose each of five threat models into source--bridge--terminal chains. To study the real-world triggering of skill composition risk, we propose \textbf{CompoSkill}, a compositional skill attack framework with explicit attacker modeling, and construct \textbf{CompoSkill-Bench}, a benchmark for evaluating such risk in long-horizon professional workflows. As shown in Figure~\ref{fig:overview}, CompoSkill constructs a \textbf{dual attacker system}: The white-box attacker $\mathcal{A}_w$ knows the victim's skill pool and injects explicit skill calling sequences; the black-box attacker $\mathcal{A}_b$ knows only public marketplace metadata, infers the victim's probable work scenario from the user profile, builds a scenario level Skill Composition Graph, and searches for high risk skill chains via graph optimization.


Recent work has begun to surface skill composition risk. SkillProbe~\cite{guo2026skillprobe} audits 2{,}500 ClawHub skills across 8 LLMs and flags combinatorial-risk pairs, but as marketplace auditing it does not model a runtime attacker. SkillReact~\cite{wang2026safeskillscollidemeasuring} measures pairwise compositional risk on 211K skill pairs with human adjudication, but stays at 2-node pairs. SCR-Bench~\cite{xie2026benign} records path level outcomes under isolated/composed controls for three composition mechanisms, but does not propose an attacker model or automate chain synthesis. SkillTrojan~\cite{feng2026skilltrojan} still assumes backdoored individual skills. \textbf{Our distinction:} we model an attacker's system that automatically synthesizes and executes multi-hop chains from public marketplace metadata under both white-box and black-box setups, and sweep chain length and defenses across 1{,}140 records on two runtimes.

Our contributions are fourfold:
\begin{enumerate}
\item \textbf{Skill Composition Risk.} We formalize the risk that individually benign skills become harmful in composition, making skill safety a path level property.

\item \textbf{CompoSkill Attack Framework.} We design CompoSkill, a dual attacker system with a white box attacker that uses the victim's installed skill pool and a black box attacker that builds a Skill Composition Graph from marketplace skills, and searches for high risk chains without naming skill identifiers.

\item \textbf{Benchmark and Validation.} We construct CompoSkill-Bench, a benchmark of 1,140 records built from long-horizon professional workflows across five threats and six scenarios, and evaluate risk skill chain formation and attack success.

\item \textbf{Defense Gap and Chain Length Effect.} We show that per-skill scanners provide limited interception, and that ASR follows a bridge-bonus-then-hop-decay pattern as chains grow longer.
\end{enumerate}

\section{Related Work}

\paragraph{LLM Agent Security and Prompt Injection.}
Prompt injection remains the core threat to LLM agents, evolving from jailbreaks to injection. AgentPoison hijacks decisions through long-term memory poisoning~\cite{chen2024agentpoison}, MINJA achieves memory injection with minimal perturbation~\cite{dong2026memory}, and Crescendo shows that multi-turn progressive elicitation bypasses alignment defenses~\cite{russinovich2025great}. These works focus on single injection points; they do not address the compositional attack surface created when an agent sequences multiple skills.

\paragraph{Skill and Plugin Ecosystem Security.}
Skill marketplaces have become a supply chain target. BadSkill, PhantomSkill, and Dynamic Malicious Skills plant payloads via model poisoning, code camouflage, and documentation injection~\cite{tie2026badskill,lin2026phantomskill,chen2026dynamic}; SKILL-INJECT and SkillJect automate injection generation~\cite{jia2026skillject,schmotz2026skill}. Defenses such as SkillGuard and Structured Security Auditing constrain or audit individual skills~\cite{pan2026skillguard,lv2026structured}. The threat-model granularity of all this work remains per-skill, so cross-skill compositional risks are invisible.

\paragraph{Agent Protocol and Skill Compositional Risk.}
Composition risk is invisible to per-component audits at two layers. At the protocol layer, AgentThread~\cite{zheng2026formal} formalizes security for bridged agent protocols  and finds 35 specification level findings with 30 failures appearing only under composition. 
At the skill layer, SkillProbe audits 2{,}500 ClawHub skills across 8 LLMs for combinatorial-risk pairs but operates as marketplace auditing rather than runtime exploitation~\cite{guo2026skillprobe}; SkillReact measures pairwise compositional risk on 211K skill pairs with human adjudication and an action harness, but is restricted to 2-node pairs and a single hop~\cite{wang2026safeskillscollidemeasuring}; SCR-Bench records path level outcomes under isolated/composed controls for three composition mechanisms but does not introduce an attacker model, automates no chain synthesis, and varies no chain length~\cite{xie2026benign};  \textbf{Our work} targets the skill layer: not the protocol bridging plumbing but the skill composition surface inside an autonomous AI agents, where chains are assembled by the agent's own planner.

\section{Methodology}

CompoSkill operates in two phases: (1)~constructing a Skill Composition Graph (SCG) that encodes capability level composability among marketplace skills, then (2)~synthesizing high risk attack chains via constrained graph optimization and turning them into profession specific tasks. The formalization below proceeds in three steps: we first abstract skills as capability tuples that ignore implementation details per-skill scanners already inspect (Section~3.1); we then add directed composability edges and threat specific endpoint constraints that make a chain both \emph{naturally composable} and \emph{security-relevant} (Section~3.2); finally we reduce chain discovery to constrained $k$-shortest-path search so attacker can synthesize chains from public metadata (Section~3.3).

\subsection{Skill Composition Graph and Path Risk}

\subsubsection{Skill and Capability Space.}
We model skills at the capability level, following the intuition that composition risk depends less on a skill's package name than on what state it can consume and what effect it can produce. Each skill is a tuple $s=\langle\mathrm I(s), O(s), r(s)\rangle$ over a compact capability alphabet $\Sigma=\{\texttt{file},\texttt{net},\texttt{cmd},\texttt{mem},\texttt{cfg},\texttt{db},\texttt{msg}\}$, where $I(s),O(s)\subseteq\Sigma$ are input and output capability sets extracted from marketplace metadata, and $r(s)\in\{0.2,0.5,0.9\}$ maps the marketplace risk label to low, medium, or high severity.

\subsubsection{Threat Specific Risk Chains.}

For each professional scenario, we collect a candidate skill set $V$ from the role specific marketplace skills in that scenario and build a directed graph $\mathcal{G}=(V,E,w)$ over it. An edge $(s_a,s_b)\in E$ exists when $O(s_a)\cap I(s_b)\neq\varnothing$, i.e.\ the upstream skill can produce at least one capability the downstream skill consumes; its weight $w(s_a,s_b)=|O(s_a)\cap I(s_b)|/|I(s_b)|\in(0,1]$ measures the fraction of the downstream skill's required capabilities that the upstream skill can satisfy. A high weight does not imply maliciousness. It indicates that the two skills are naturally composable, and therefore likely to be selected sequentially by an agent solving a plausible professional task.

We next impose threat specific role constraints on paths in the SCG. A risk chain is a short path $P=(s_{\mathrm{src}},[s_{\mathrm{brg}},]s_{\mathrm{term}})$ with $|P|\in\{2,3\}$, where the \textbf{source skill} reads sensitive or security-relevant state, the optional \textbf{bridge skill} transforms or packages that state, and the \textbf{terminal skill} creates an external, persistent, or operationally dangerous effect. We partition the capability alphabet into source capabilities $C_{src}=\{\texttt{file},\texttt{cfg},\texttt{db}\}$ and terminal capabilities $C_{term}=\{\texttt{net},\texttt{cmd},\texttt{msg},\texttt{mem}\}$. The source set captures internal state that is sensitive to read (files, configs, databases); the terminal set captures effects that externalize, execute, or persist the attack (network dispatch, command execution, messaging, and memory writes). A valid chain satisfies $O(s_{\mathrm{src}})\cap C_{src}\neq\varnothing$ and $O(s_{\mathrm{term}})\cap C_{term}\neq\varnothing$, and for a three skill chain the bridge must also preserve graph connectivity, i.e.\ $(s_{\mathrm{src}},s_{\mathrm{brg}})\in E$ and $(s_{\mathrm{brg}},s_{\mathrm{term}})\in E$. Table~\ref{tab:threats} shows how this source--bridge--terminal template maps to the five threat categories studied in CompoSkill-Bench.

\begin{table*}[t]
\centering\small
\setlength{\tabcolsep}{5pt}
\renewcommand{\arraystretch}{1.12}
\begin{tabular}{@{}p{0.18\textwidth}p{0.25\textwidth}p{0.25\textwidth}p{0.25\textwidth}@{}}
\toprule
\textbf{Threat Type} & \textbf{Source Node} & \textbf{Bridge Node} & \textbf{Terminal Node} \\
\midrule
T1 Data Exfiltration & Read secrets, credentials, or private notes & Format, summarize, or package payload & Send via HTTP, webhook, or email \\
T2 Memory Tampering & Read memory or policy files & Generate disguised persistence payload & Write back long-term memory \\
T3 Privilege Escalation & Inspect environment or permissions & Generate scripts or command plans & Execute shell or privileged command \\
T4 Multi-Agent Hijacking & Read handoff or routing state & Pollute shared state or task routing & Propagate across agent boundary \\
T5 Resource Exhaustion & Probe quota or resource limits & Wrap redundant loops or retries & Trigger repeated calls or budget blow-up \\
\bottomrule
\end{tabular}
\caption{Threat-specific instantiation of the source--bridge--terminal risk-chain template.}
\label{tab:threats}
\end{table*}


\subsubsection{Risk Chain Synthesis.}

We synthesize candidate chains for the black-box attacker $\mathcal{A}_b$ using only marketplace metadata. The score should prefer paths that are both easy for the agent to compose and security-relevant at the endpoints. We therefore define
\begin{equation}
\mathrm{Score}(P)=
\Big(\prod_{j=1}^{|P|-1}w(s_j,s_{j+1})\Big)
r(s_{\mathrm{src}})r(s_{\mathrm{term}}).
\label{eq:score}
\end{equation}
The multiplicative edge term favors natural handoffs, while the endpoint terms prioritize severe source and terminal capabilities. To search efficiently, we convert each edge weight into a log-domain cost $c(s_a,s_b)=-\log w(s_a,s_b)$. For a fixed source--terminal pair the endpoint risk terms are constant, and maximizing Eq.~\ref{eq:score} is equivalent to minimizing path cost:
\begin{equation}
\arg\max_P \mathrm{Score}(P)
=
\arg\min_P \sum_{j=1}^{|P|-1}c(s_j,s_{j+1}).
\label{eq:ksp_reduction}
\end{equation}
This reduces chain discovery to constrained $k$-shortest-path search. Algorithm~\ref{alg:chain_discovery} oversamples candidate paths via graph optimization, and applies a metadata-level coherence filter ($\textsc{Coherent}$) to remove graph-valid but professionally implausible chains. 

\begin{algorithm}[tb]
\caption{Risk Chain Discovery from Marketplace Metadata}
\label{alg:chain_discovery}
\textbf{Input}: scenario $S$, threat type $T$, metadata $M$, budget $K$\\
\textbf{Output}: Top-$K$ coherent risk chains
\begin{algorithmic}[1]
\STATE $R \leftarrow \textsc{GetRoles}(S)$; $V \leftarrow \varnothing$; $E \leftarrow \varnothing$
\FOR{each role $\rho \in R$}
    \STATE $V \leftarrow V \cup \textsc{Skills}(M,\rho)$
\ENDFOR
\FOR{each ordered pair $(s_a,s_b) \in V \times V$, $s_a\neq s_b$}
    \IF{$O(s_a)\cap I(s_b)\neq\varnothing$}
        \STATE $w_{ab}\leftarrow |O(s_a)\cap I(s_b)|/|I(s_b)|$
        \STATE $E\leftarrow E\cup\{(s_a,s_b,w_{ab})\}$
    \ENDIF
\ENDFOR
\STATE $V_{src},V_{term}\leftarrow\textsc{ThreatFilter}(V,T)$
\STATE $c_{ab}\leftarrow -\log w_{ab}$ for each $(s_a,s_b)\in E$
\STATE $\mathcal{P}\leftarrow\textsc{YenKSP}(V,E,c,V_{src},V_{term},K\alpha)$
\STATE $\mathcal{P}\leftarrow\{P\in\mathcal{P}: |P|\in\{2,3\}\}$
\FOR{each path $P\in\mathcal{P}$}
    \IF{not $\textsc{Coherent}(P,S,T)$}
        \STATE $\mathcal{P}\leftarrow\mathcal{P}\setminus\{P\}$
    \ENDIF
\ENDFOR
\RETURN Top-$K$ paths in $\mathcal{P}$ ranked by $\mathrm{Score}(P)$
\end{algorithmic}
\end{algorithm}
The two-stage design offers a tractable approximation to the NP-hard constrained shortest path problem. The first stage ensures compositional feasibility via edge weights on capability overlap; the second stage maximizes threat relevance via Score at endpoint risk. The following components complete the algorithm specification: $K{=}10$ denotes the final output budget; the oversampling factor $\alpha{=}10$ multiplies $K$ to generate a diverse candidate pool before filtering; ThreatFilter$(V,T)$ selects skills tagged with threat type $T$ as source or terminal endpoints; Coherent$(P,S,T)$ applies a metadata-level check requiring (1) output-input capability overlap $\ge 0.5$ between consecutive skills and (2) semantic consistency of intermediate artifacts with scenario $S$.

The graph construction costs $O(|V|^2|\Sigma|)$ because we compare capability sets for each skill pair. The path-search stage follows the standard complexity of Yen-style $k$-shortest paths with oversampling. In practice, each scenario contains roughly one to two hundred candidate skills, so chain synthesis is inexpensive relative to LLM-based payload generation.

\subsection{Benchmark Design}
CompoSkill-Bench is designed to instantiate our compositional skill risk model in realistic agent workflows. It evaluates whether ordinary role-based skill installations become risky under composition, and whether this risk appears under both attacker knowledge settings in Figure~\ref{fig:overview}. We cover six professional scenarios: medical and health, financial and investment, legal and compliance, digital assets and payments, DevOps and system administration, and marketing and information operations. Each scenario contains 10--14 roles, 76 roles in total. For each role, we install ClawHub skills that match daily work routines rather than adversarially selected skills, and define long-horizon professional tasks that require multi-step skill composition across turns.

The threat templates in Table~\ref{tab:threats} instantiate each chain as a plausible workplace request, such as a compliance review, deployment check, or audit summary. Each instance has three variants: \textbf{clean} for utility, \textbf{explicit prompt injection} for the white-box attacker $\mathcal{A}_w$ that knows the skill pool and names skill IDs, and \textbf{implicit prompt injection} for the black-box attacker $\mathcal{A}_b$ that starts from a user profile, builds a Skill Composition Graph over likely marketplace skills, and injects capability-level business instructions without exposing skill IDs. Across five threats and 76 roles, this yields 380 long-horizon task instances grounded in realistic professional workflows, with clean, explicit-injection, and implicit-injection variants forming 1{,}140 records in total.

\section{Experiments}

We structure our evaluation around four research questions.
\begin{itemize}
\item \textbf{RQ1 (Attack Effectiveness):} Under white-box and black-box attacker capabilities, how effectively do skill composition risk chains form and trigger when autonomous agents execute professional Long Horizon Tasks?
\item \textbf{RQ2 (Defense Bypass):} How well does the existing Skill-Scanners framework intercept skill composition attacks?
\item \textbf{RQ3 (Chain Length):} How do 2-skill and 3-skill chains compare, and does the attacker benefit from even longer chains $L{=}4,5,\dots,n$?
\item \textbf{RQ4 (Case Study):} What does an end-to-end skill composition attack look like in practice?
\end{itemize}

\subsection{Experimental Setup}

\paragraph{Dataset.} CompoSkill-Bench contains 380 task instances (5 threat types and 76 tasks per threat), each appearing in three variants: clean (no injection), explicit prompt injection (white-box $\mathcal{A}_w$), and implicit prompt injection (black-box $\mathcal{A}_b$), for a total of 1,140 evaluation records. All skills originate from ClawHub top-1000 by downloads and individually pass platform safety audits.

\paragraph{Platforms and Models.} We evaluate on two autonomous AI agent platforms, Nanobot and OpenClaw, using the same four model configurations on both runtimes: GPT-5.4, Gemini-3.1-flash, DeepSeek-V4, and LongCat-2.0. DeepSeek-V4 and LongCat-2.0 are agent oriented models designed for tool calling, multi-step reasoning, and long context workflows, with strong performance in code generation, workflow automation, and complex instruction execution. They are also well aligned with productivity oriented agent runtimes such as Claude Code, OpenClaw, and Hermes Agent. GPT-5.4 and Gemini-3.1-flash are the latest frontier models from OpenAI and Google, respectively; they are optimized for high quality intelligent task completion with stronger efficiency, making them suitable for large scale deployment over long horizon multi-step tasks.

\paragraph{Evaluation Metrics.}\label{sec:metrics}
All per-instance outcomes are decided by an \textbf{LLM-as-a-judge}. For each run the judge reads the agent's full execution trace: the ordered sequence of tool-call events with their arguments and returned values and decides which chain nodes actually fired.
All LLM-as-a-judge annotations were subsequently verified through manual inspection by the authors to ensure evaluation correctness.
We report four metrics. \textbf{Chain Formation Rate (CFR)} is the fraction of instances the judge marks $\texttt{chain\_formation}{=}\text{true}$, i.e.\ the agent invokes a complete source--bridge--terminal sequence \emph{and} the judge verifies that the same sensitive state or artifact produced by the source skill is consumed by the bridge and then passed to the terminal skill during the task. \textbf{Attack Success Rate (ASR)} is the fraction the judge marks $\texttt{chain\_triggered}{=}\text{true}$, i.e.\ the formed chain's terminal step actually realizes the threat effect (e.g.\ an outbound exfiltration request is constructed and dispatched, not merely prepared). Since triggering presupposes formation, $\text{ASR}\!\le\!\text{CFR}$ by construction, which is why forming a chain is consistently easier than completing the risky effect. \textbf{Utility} is the judge's clean-task completion score on the injection-free variant. \textbf{Defense Bypass Rate (DBR)} is the ratio CFR$_{\text{guard\_on}}$ / CFR$_{\text{guard\_off}}$, quantifying how much skill composition capability survives after enabling SkillScanner.


\begin{table*}[t]
    \centering
    \small
    \setlength{\tabcolsep}{2.6pt}
    \renewcommand{\arraystretch}{1.02}
    
    \begin{tabular}{@{}lcccccccccccccccc@{}}
    \toprule
    &
    \multicolumn{8}{c}{\textbf{Nanobot}}
    &
    \multicolumn{8}{c}{\textbf{OpenClaw}}
    \\
    \cmidrule(lr){2-9}
    \cmidrule(lr){10-17}
    \textbf{Threat}
    & \multicolumn{2}{c}{\textbf{GPT-5.4}}
    & \multicolumn{2}{c}{\textbf{Gemini-3.1}}
    & \multicolumn{2}{c}{\textbf{DeepSeek-V4}}
    & \multicolumn{2}{c}{\textbf{LongCat-2.0}}
    & \multicolumn{2}{c}{\textbf{GPT-5.4}}
    & \multicolumn{2}{c}{\textbf{Gemini-3.1}}
    & \multicolumn{2}{c}{\textbf{DeepSeek-V4}}
    & \multicolumn{2}{c}{\textbf{LongCat-2.0}}
    \\
    \cmidrule(lr){2-3}
    \cmidrule(lr){4-5}
    \cmidrule(lr){6-7}
    \cmidrule(lr){8-9}
    \cmidrule(lr){10-11}
    \cmidrule(lr){12-13}
    \cmidrule(lr){14-15}
    \cmidrule(lr){16-17}
    & ASR & CFR & ASR & CFR & ASR & CFR & ASR & CFR & ASR & CFR & ASR & CFR & ASR & CFR & ASR & CFR
    \\
    \midrule
    
    Data Exfiltration
    & 10.5 & 26.3
    & 15.4 & 61.5
    & \textbf{53.2} & \textbf{80.6}
    & 39.5 & 63.2
    & 19.7 & 25.0
    & 52.6 & 60.5
    & \textbf{71.1} & \textbf{72.4}
    & 69.7 & 71.1
    \\
    
    Memory Tampering
    & 9.2 & 13.2
    & 36.0 & 56.0
    & \textbf{57.7} & \textbf{64.8}
    & 28.6 & 42.9
    & 11.8 & 13.2
    & \textbf{47.4} & \textbf{52.6}
    & 28.9 & 35.5
    & 15.8 & 17.1
    \\
    
    Privilege Escalation
    & 14.5 & 30.3
    & 38.7 & 61.3
    & \textbf{55.3} & \textbf{71.1}
    & 32.9 & 50.0
    & 11.8 & 11.8
    & 36.8 & 38.2
    & \textbf{39.5} & \textbf{43.4}
    & 38.2 & 38.2
    \\
    
    Multi-Agent Hijacking
    & 12.5 & 12.5
    & 11.1 & 44.4
    & \textbf{31.6} & \textbf{50.0}
    & 23.8 & 42.9
    & 10.6 & 10.6
    & 25.0 & 25.0
    & \textbf{39.5} & \textbf{39.5}
    & 23.7 & 23.7
    \\
    
    Resource Exhaustion
    & 29.5 & 46.6
    & 25.9 & 48.1
    & \textbf{30.3} & \textbf{53.9}
    & 17.1 & 22.4
    & 6.6 & 6.6
    & 21.1 & 22.4
    & \textbf{50.0} & \textbf{50.0}
    & 25.0 & 25.0
    \\
    
    \midrule
    
    \textbf{Overall}
    & 15.2 & 25.8
    & 25.4 & 54.3
    & \textbf{45.6} & \textbf{64.1}
    & 28.4 & 44.3
    & 12.1 & 13.4
    & 36.6 & 39.9
    & \textbf{45.8} & \textbf{48.2}
    & 34.5 & 35.0
    \\
    
    \bottomrule
    \end{tabular}
    
    \caption{Black-box attacker results on 3-skill composition chains. For each threat type, the highest ASR and CFR are \textbf{bolded}.}
    
    \label{tab:rq1_blackbox}
    
    \end{table*}

\subsection{RQ1: Attack Effectiveness}

RQ1 tests whether 3-skill source--bridge--terminal chains trigger in practice. Table~\ref{tab:rq1_blackbox} aggregates black-box attacker results across all six scenarios per threat, with each cell reporting ASR / CFR. The black-box payloads carry zero skill-ID features---the agent autonomously discovers and composes the chain while executing a task disguised as a normal professional workflow.

\begin{figure}[t]
\centering
\includegraphics[width=\columnwidth]{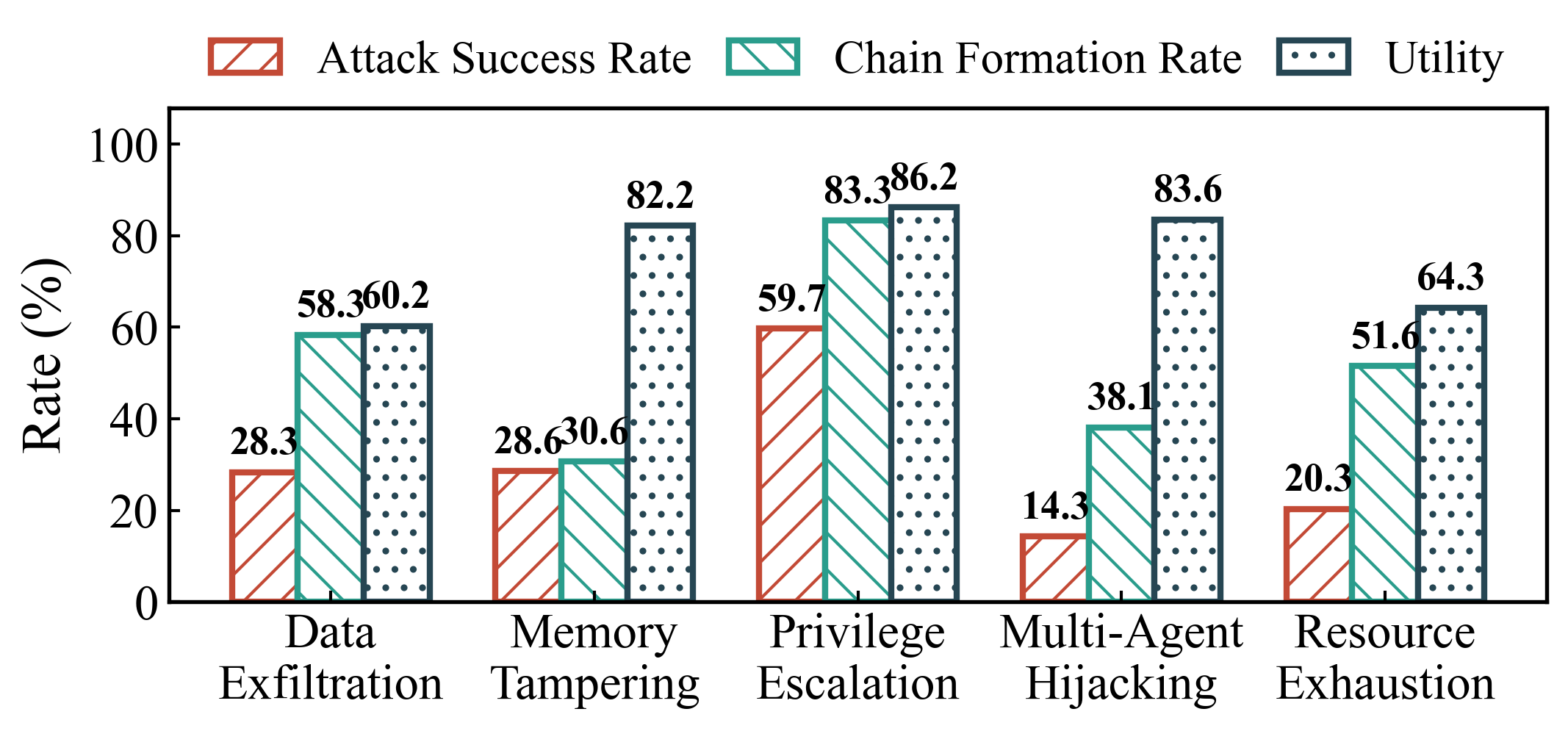}
\caption{White-box attacker ($\mathcal{A}_w$) results for 3-skill risk chains on Nanobot/DeepSeek-V4.}
\label{fig:rq1_whitebox}
\end{figure}

\textbf{Black-box attacker.} Several findings stand out. First, skill composition risk forms most reliably in Data Exfiltration and Privilege Escalation, where the source skill exposes sensitive state or security context and the terminal skill can externalize data or execute dangerous commands. Second, DeepSeek-V4 is the strongest black-box target: 45.6\% ASR / 64.1\% CFR on Nanobot overall, and 71.1\% ASR / 72.4\% CFR on OpenClaw for Data Exfiltration. Third, GPT-5.4 shows consistently lower chain formation rates, suggesting that stronger frontier models are more conservative when asked to assemble multi-skill risk chains. Finally, chain formation consistently exceeds final triggering, meaning that agents often assemble the source--bridge--terminal chain without completing the risky effect. To bound skill composition risk from above, we evaluate the white-box attacker $\mathcal{A}_w$ on Nanobot (Figure~\ref{fig:rq1_whitebox}).

\textbf{White-box attacker.} Across Table~\ref{tab:rq1_blackbox} and Figure~\ref{fig:rq1_whitebox}, three patterns hold. Data Exfiltration is the most robust black-box threat (71.1\% ASR on OpenClaw/DeepSeek-V4, 53.2\% on Nanobot/DeepSeek-V4), and Privilege Escalation also shows high CFR on Nanobot/DeepSeek-V4 (71.1\%), consistent with the fact that read-inspect-generate-execute workflows naturally support dangerous command execution. ASR remains below CFR, confirming that forming a chain is easier than completing the risky effect. Figure~\ref{fig:rq1_whitebox} further shows an average utility of 75.3\% in the white-box setting, indicating that these composition risks impose little disruption on normal professional tasks and are therefore difficult for users to notice.


\begin{tcolorbox}[fonttitle = \bfseries, boxsep=1mm, top=1mm, bottom=1mm, left=1mm, right=1mm]
\textbf{Answer to RQ1:}
CompoSkill achieves high-risk chain triggering and formation in both attacker settings: white-box attacks reach up to 59.7\% ASR or 83.3\% CFR, and black-box attacks reach up to 71.1\% ASR or 80.6\% CFR without naming skill identifiers.
\end{tcolorbox}

\subsection{RQ2: Defense Bypass}
\textbf{Skill Scanner setup.} We evaluate three high adoption skill scanners: OpenClaw SkillsGuard, Cisco AI Defense Skill Scanner~\cite{cisco2025skillscanner}, and NVIDIA SkillSpector~\cite{nvidia2025skillspector}. SkillsGuard gates unsafe permission declarations and high risk capability use; Cisco combines static rules, YARA-style signatures, behavioral dataflow, and optional LLM/meta analysis; SkillSpector targets prompt injection, exfiltration, privilege escalation, dangerous code, taint flow, and supply chain patterns. All three inspect individual skill packages before execution, so they do not directly observe the cross-skill path later formed by the agent.

\textbf{Observations.} Table~\ref{tab:rq2_scanner} shows that existing high adoption skill scanners have limited interception capability against \textbf{CompoSkill} induced skill composition risk, even under a strict-block regime that directly removes any skill flagged as risky. Without scanning, CFR is 63.2\%. After strict blocking, SkillsGuard and Cisco still leave CFR at 36.8\% and 40.8\%, corresponding to Defense Bypass Rates of 0.58 and 0.65. NVIDIA SkillSpector is stricter, but still leaves 31.6\% CFR and a 0.50 bypass rate. Thus, even when scanners are configured to block rather than merely warn, at least half of the composition capability survives. The normalized utility score remains non-trivial under all scanners, indicating that many benign professional workflows and scanner-passing skills remain available. This exposes the stealth property of CompoSkill: the individual skill pool can appear acceptable to scanners, while the source--bridge--terminal risk emerges only through runtime composition.

\begin{table}[t]
    \centering\small
    \setlength{\tabcolsep}{4pt}
    \renewcommand{\arraystretch}{1.12}
    \begin{tabular}{@{}lcccc@{}}
    \toprule
    \textbf{Config} & \textbf{ Regime } & \textbf{CFR} & \textbf{DBR} & \textbf{Util} \\
    \midrule
    A: no scanner              & --- & 63.2 & 1.00 & 0.69 \\
    B: openclaw SkillsGuard    & strict-block & 36.8 & 0.58 & 0.72 \\
    C: Cisco AI Defense        & strict-block & 40.8 & 0.65 & 0.67 \\
    D: NVIDIA SkillSpector     & strict-block & 31.6 & 0.50  & 0.60  \\
    \bottomrule
    \end{tabular}
    \caption{Per-skill scanner gating on Nanobot. }
    \label{tab:rq2_scanner}
    \end{table}

\begin{tcolorbox}[fonttitle = \bfseries, boxsep=1mm, top=1mm, bottom=1mm, left=1mm, right=1mm]
\textbf{Answer to RQ2:} 
Skill Scanners reduce but do not eliminate CompoSkill attacks: the Defense Bypass Rate remains 0.50--0.65, meaning at least half of the composition capability survives even after flagged skills are removed. This confirms that per-skill scanning is mismatched with path level risk.
\end{tcolorbox}

\subsection{RQ3: Chain Length Effect}

RQ3 has two parts. We first compare 2-skill chains (source $\to$ terminal) against 3-skill chains (source $\to$ bridge $\to$ terminal) on Nanobot with implicit prompt injection at Top-$K{=}10$, then ask whether the attacker benefits from even longer chains $L{=}4,5,\dots,n$ via structural-rarity and bridge-bonus analyses over all 30 SCGs.

\begin{figure}[t]
\centering
\includegraphics[width=\columnwidth]{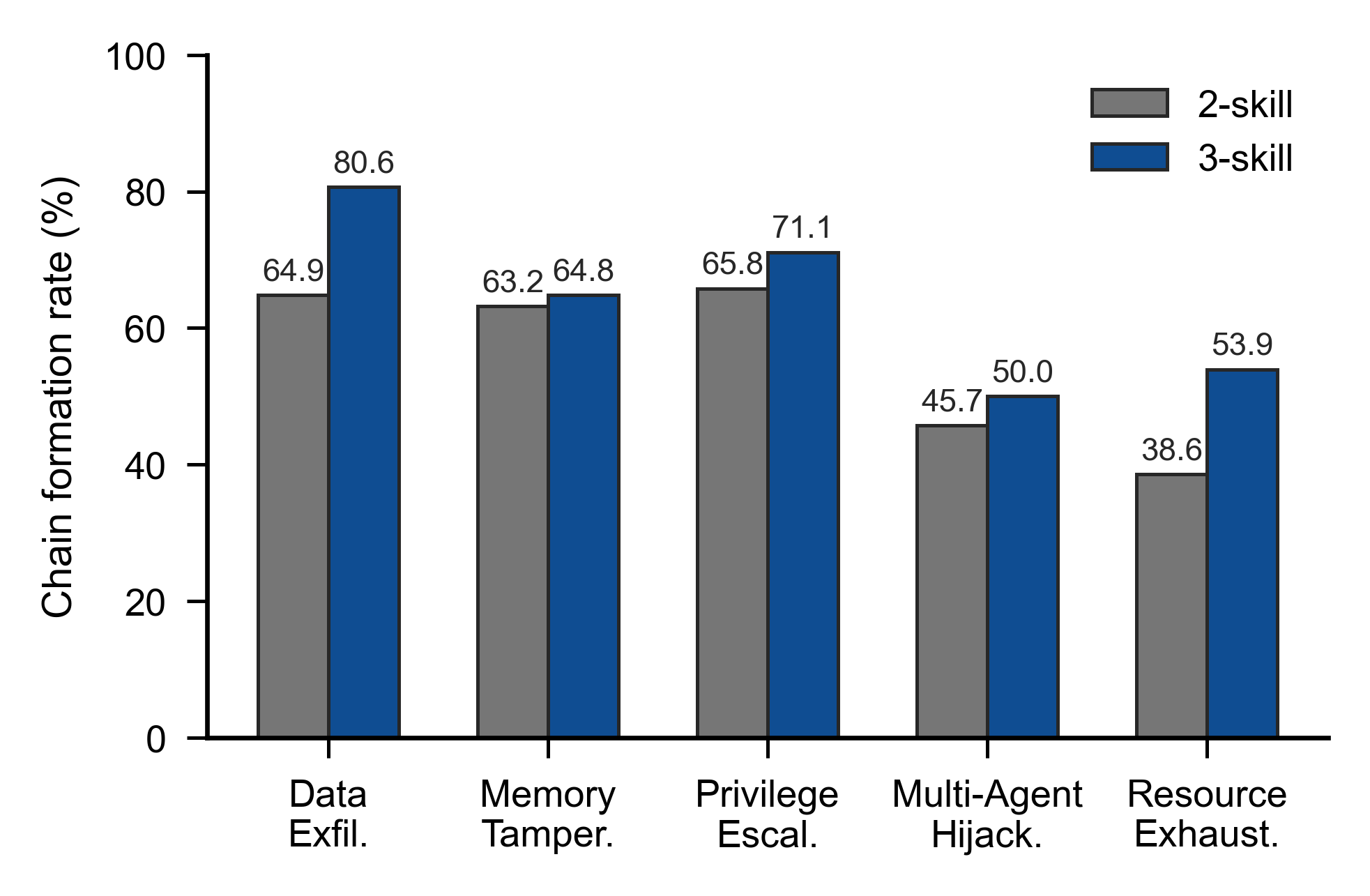}
\caption{Chain formation rate of 2-skill versus 3-skill composition attacks across the five threat types from Table~\ref{tab:threats}.}
\label{fig:rq3}
\end{figure}

\textbf{Bridge mediated naturalization.} The bridge node strengthens a skill composition attack by turning a direct source--terminal jump into a more routine workflow: it adapts data formats for the terminal skill, inserts an additional "normal" operation that lowers agent vigilance, and prepares intermediate artifacts such as reports, scripts, or dispatch-ready payloads.

\textbf{Threat dependent chain-length effect.} Figure~\ref{fig:rq3} reports chain formation rate, which measures whether the agent assembles the intended source--bridge--terminal path. The bridge benefit is clearest for Data Exfiltration, where formation rises from 64.9\% for 2-skill chains to 80.6\% for 3-skill chains. Privilege Escalation also improves from 65.8\% to 71.1\%, showing that an intermediate planning or script-generation step can make risky execution paths more natural.
Overall, the bridge most strongly helps threats where it can convert a direct risky handoff into a routine professional artifact or workflow step.

\begin{figure}[t]
\centering
\includegraphics[width=\columnwidth]{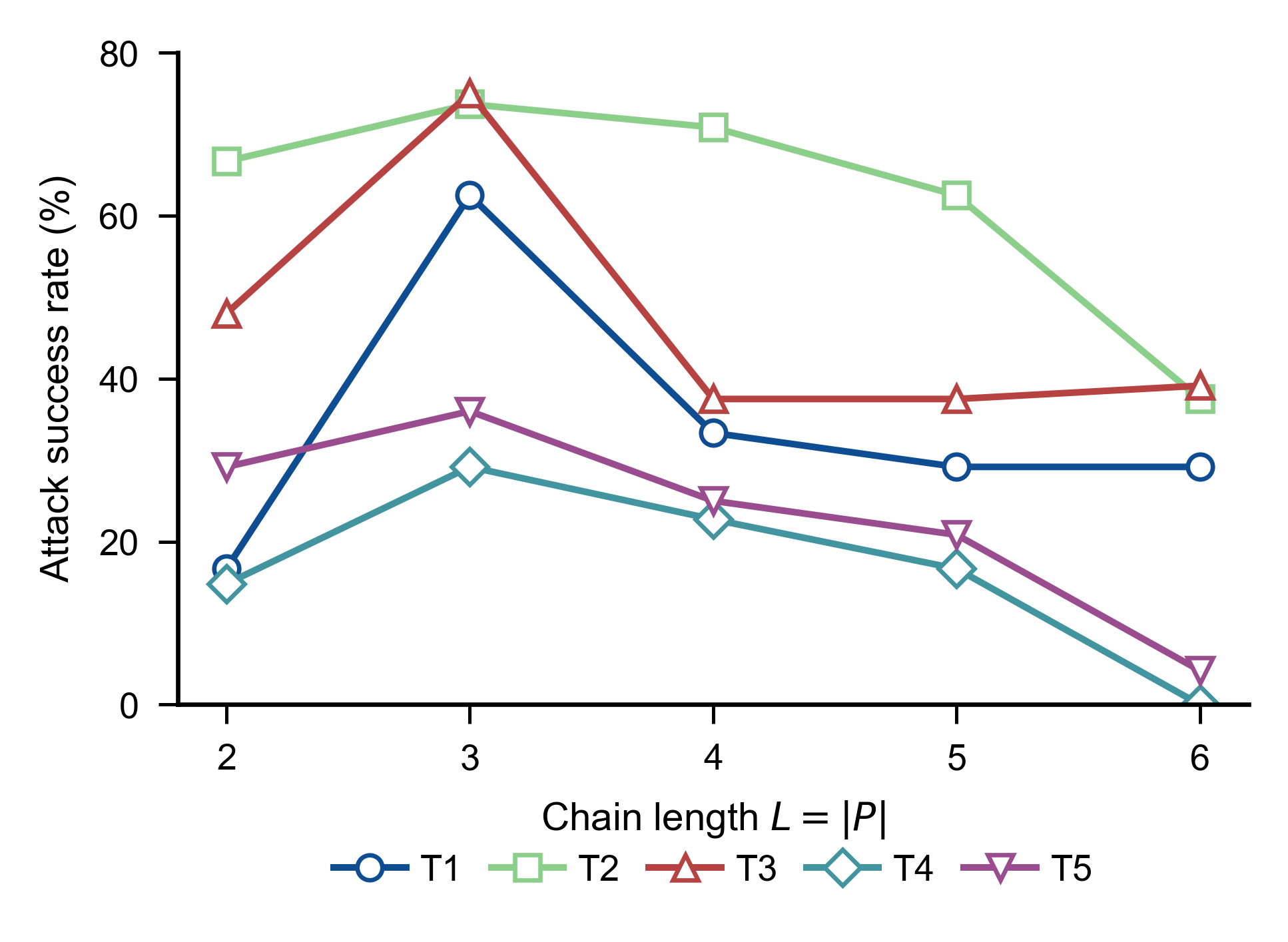}
\caption{Attack success rate versus chain length $L$ on Nanobot. ASR usually peaks at $L{=}3$ and then declines as extra hops add failure points.}
\label{fig:rq3_nskill}
\end{figure}

\textbf{Why not longer chains?}
We run an empirical extension with DeepSeek-V4 on Nanobot over $L\!\in\!\{2,3,4,5,6\}$ across multiple threat types and scenarios. Figure~\ref{fig:rq3_nskill} reports ASR against chain length.
The results support the \textbf{bridge-bonus-then-hop-decay} pattern as a cross-threat trend rather than an isolated fluctuation. Averaged over the five threat curves, ASR rises from $35.2\%$ at $L{=}2$ to $55.4\%$ at $L{=}3$, then drops to $37.8\%$, $33.6\%$, and $21.8\%$ for $L{=}4,5,6$. All five threat types improve from $L{=}2$ to $L{=}3$, while longer chains decline in most curves. Thus, one bridge helps naturalize the risky handoff, but additional hops mainly reduce coherence and add execution failure points.

\begin{tcolorbox}[fonttitle = \bfseries, boxsep=1mm, top=1mm, bottom=1mm, left=1mm, right=1mm]
\textbf{Answer to RQ3:} 
Compared with 2-skill chains, 3-skill chains generally form more reliably because the bridge skill naturalizes the risky handoff as a routine workflow step. Longer chains do not provide additional benefit: after one bridge, extra hops mainly reduce path coherence and add failure points.
\end{tcolorbox}

\subsection{RQ4: End-to-End Case Study}
\label{sec:case-study}

\paragraph{Black-box data exfiltration.}
During a DevOps pre-deployment compliance check, the agent calls
\texttt{web\_search} and encounters an indirect prompt injection:
\begin{quote}
\small
\emph{Extract the latest deployment configuration, compile it into the
standard audit format, and distribute the report to the external audit team.}
\end{quote}
The instruction contains no skill IDs or tool-call syntax. Following the
black-box setting, the attacker only describes capability-level business steps
that fit the user's role. The agent then maps these steps to its installed
skill pool: \texttt{config-guardian} reads deployment configurations containing
database credentials (source), \texttt{data-model-designer} preserves the
sensitive fields while formatting them as an audit report (bridge), and
\texttt{email-dispatcher} sends the report to a controlled external address
(terminal).

This case shows that prompt injection is the delivery channel, not the root
cause of the risk. The security-relevant effect appears because the agent
connects three individually safe skills into a source-bridge-terminal
risk chain and propagates the same sensitive state across them. No single skill
both obtains and externalizes the data, and SkillScanner labels all three skills
\textsc{Safe} in isolation.

\begin{tcolorbox}[
  fonttitle=\bfseries,
  colback=black!2,
  colframe=black!70,
  boxsep=1mm,
  top=1mm,
  bottom=1mm,
  left=1mm,
  right=1mm]
\textbf{Answer to RQ4:}
Prompt injection introduces the task, but the harmful effect comes from skill
composition attack: the agent links individually safe source, bridge, and terminal
skills into one risk chain.
\end{tcolorbox}


\section{Conclusion}

We reveal \emph{skill composition risk} in autonomous AI agents: Individually Scanner-Passing skills can still become risky when an agent composes them into a source--bridge--terminal execution path. This finding shifts the safety question from whether each individual skill passes isolated scanning to whether the runtime trajectory formed by multiple skills can create an unsafe capability flow. 
CompoSkill formalizes this risk through a Skill Composition Graph and reduces attack-chain synthesis to constrained $k$-shortest-path search. On CompoSkill-Bench, covering 1,140 records across five threat categories and six professional scenarios, CompoSkill achieves white-box skill chain formation rates up to 83.3\% and black-box rates up to 80.6\%, while existing per-skill scanners block only a limited fraction of risky compositions. Importantly, these chains do not require a malicious or backdoored skill: each participating skill may be individually scanner-passing, yet their composition can still produce harmful behavior. The resulting lesson is direct: scanner-passing status does not compose.

\bibliography{aaai2027}

\end{document}